\documentclass[cits]{PoS}
\usepackage{clshan-math,clshan-dm-dd}

\def\gsim{\:\raisebox{-0.5ex}{$\stackrel{\textstyle>}{\sim}$}\:}
\title{Constraining the Spin--Independent WIMP--Nucleon Coupling
       from Direct Dark Matter Detection Data}

\ShortTitle{Constraining the SI WIMP--nucleon coupling from direct DM
  detection data} 

\author{Manuel Drees \\
     Physikalisches Institut der Universit\"at Bonn, D--53115 Bonn, Germany \\
        School of Physics, KIAS, Seoul 130--012, Republic of Korea  \\
    Bethe Center of Theoretical Physics, Universit\"at Bonn, D--53115 Bonn,
     Germany \\ 
        E-mail: \email{drees@th.physik.uni-bonn.de}}

\author{\speaker{Chung-Lin Shan} \\ 
        School of Physics and Astronomy, Seoul National University,
        Seoul 151--747, Republic of Korea \\
        E-mail: \email{cshan@hep1.snu.ac.kr}}
      
      \abstract{ Weakly Interacting Massive Particles (WIMPs) are one of the
        leading candidates for Dark Matter.  For understanding the properties
        of WIMPs and identifying them among new particles produced at
        colliders (hopefully in the near future), determinations of their mass
        and their couplings on nucleons from direct Dark Matter detection
        experiments are essential.  Based on our method for determining the
        WIMP mass model--independently from experimental data, we present a
        way to also estimate the spin--independent (SI) WIMP--nucleon coupling
        by using measured recoil energies directly.  This method is
        independent of the as yet unknown velocity distribution of halo WIMPs.
        In spite of the uncertainty of the local WIMP density (of a factor of
        $\sim$ 2), at least an upper limit on the SI WIMP--nucleon coupling
        could be given, once two (or more) experiments with different target
        nuclei obtain positive signals.  In a background--free environment,
        for a WIMP mass of 100 GeV its SI coupling on nucleons could in
        principle be estimated with a statistical error of only \mbox{$\sim$
          15\%} with just 50 events from each experiment.  }

\FullConference{Identification of dark matter 2008 \\
                August 18-22, 2008 \\
                Stockholm, Sweden}

\begin{document}
\section{Introduction}

There is strong evidence that more than 80\% of all matter in the Universe is
dark (i.e., interacts at most very weakly with electromagnetic radiation and
ordinary matter). The dominant component of this cosmological Dark Matter must
be due to some yet to be discovered, non--baryonic particles.  Weakly
Interacting Massive Particles (WIMPs) $\chi$ with masses roughly between 10
GeV and a few TeV are one of the leading candidates for Dark Matter (for
reviews, see Refs.~\cite{SUSYDM}).

Currently, the most promising method to detect many different WIMP candidates
is the direct detection of the recoil energy deposited in a low--background
laboratory detector by elastic scattering of ambient WIMPs on the target
nuclei \cite{deta}.  The differential rate for elastic WIMP--nucleus
scattering is given by \cite{SUSYDM}:
\beq \label{eqn:dRdQ}
   \dRdQ
 = \calA \FQ \int_{v_{\rm min}}^{v_{\rm max}} \bfrac{f_1(v)}{v} dv\, .
\eeq
Here $R$ is the direct detection event rate, i.e., the number of events per
unit time and unit mass of detector material, $Q$ is the energy deposited in
the detector, $F(Q)$ is the elastic nuclear form factor, $f_1(v)$ is the
one--dimensional velocity distribution function of the WIMPs impinging on the
detector, $v$ is the absolute value of the WIMP velocity in the laboratory
frame.  The constant coefficient $\calA$ is defined as
\beq \label{eqn:calA}
        \calA
 \equiv \frac{\rho_0 \sigma_0}{2 \mchi m_{\rm r,N}^2}\, ,
\eeq
where $\rho_0$ is the WIMP density near the Earth and $\sigma_0$ is the total
cross section ignoring the form factor suppression. The reduced mass $m_{\rm
  r,N}$ is defined by
\beq \label{eqn:mrN}
        m_{\rm r,N}
 \equiv \frac{\mchi \mN}{\mchi+\mN}\, ,
\eeq
where $\mchi$ is the WIMP mass and $\mN$ that of the target nucleus.  Finally,
$\vmin = \alpha \sqrt{Q}$ is the minimal incoming velocity of incident WIMPs
that can deposit the energy $Q$ in the detector with
\beq \label{eqn:alpha}
        \alpha
 \equiv \sfrac{\mN}{2 m_{\rm r,N}^2}\, ,
\eeq
and ${v_{\rm max}}$ is relared to the escape velocity from our Galaxy at the
position of the Solar system.

\section{Estimating the SI WIMP--nucleon coupling}

Based on our work on the reconstruction of the (moments of the) velocity
distribution of halo WIMPs \cite{DMDDf1v}, the integral over the
one--dimensional WIMP velocity distribution on the right--hand side of
Eq.(\ref{eqn:dRdQ}), which is the minus--first moment of this distribution,
can be estimated by \cite{DMDDmchi}
\beq \label{eqn:moments}
   \expv{v^{-1}}(v(\Qmin), v(\Qmax))
 = \int_{v(\Qmin)}^{v(\Qmax)} \bfrac{f_1(v)}{v} dv
 = \frac{1}{\alpha} \bfrac{2 r(\Qmin)/\FQmin}{2 \Qmin^{1 /2} r(\Qmin)/
   \FQmin+I_0}\, . 
\eeq
Here $v(Q) = \alpha \sqrt{Q}$, $Q_{\rm (min,max)}$ are the minimal and maximal
(cut--off) energies of the experimental data set, respectively, $r(\Qmin)
\equiv (dR/dQ)_{Q = \Qmin}$, and $I_n(\Qmin,\Qmax)$ can be estimated by
\beq \label{eqn:In_sum}
   I_n(Q_{\rm min}, Q_{\rm max})
 = \sum_a \frac{Q_a^{(n-1)/2}}{F^2(Q_a)}\, ,
\eeq
where the sum runs over all events in the data set that satisfy $Q_a \in
[Q_{\rm min}, Q_{\rm max}]$.\footnote{Note that the generalized moments
  $\expv{v^n}(v(\Qmin), v(\Qmax))$ are independent of the local WIMP density,
  $\rho_0$, as well as of the WIMP--nucleus cross section, $\sigma_0$.
  Moreover, one does not need to know $f_1(v)$ in order to determine its
  moments via a generalization of Eq.(\ref{eqn:moments}).  Every term needed
  in this paper e.g., $r(\Qmin)$ and $I_n(Q_{\rm min}, Q_{\rm max})$, can be
  estimated either from a functional form of the scattering spectrum or from
  experimental data (i.e., the measured recoil energies) directly. More
  details about estimating $r(\Qmin)$, $I_n(\Qmin,\Qmax)$, their statistical
  errors and the other formulae needed can be found in Refs.~\cite{DMDDf1v},
  \cite{DMDDmchi}.} On the other hand, using the assumption that the
spin--independent (SI) WIMP scattering cross section is the same for both
protons and neutrons, the ``pointlike'' cross section $\sigma_0$ of
Eq.(\ref{eqn:calA}) can be written as
\beq \label{eqn:sigma0SI}
   \sigma_0
 = \afrac{4}{\pi} m_{\rm r,N}^2 A^2 |f_{\rm p}|^2\, ,
\eeq
where $f_{\rm p}$ is the effective $\chi \chi {\rm p p}$ four--point coupling,
and $A$ is the number of nucleons in the nucleus.

Substituting Eqs.(\ref{eqn:moments}) and (\ref{eqn:sigma0SI}) into
Eq.(\ref{eqn:dRdQ}), it can easily be found that
\beq \label{eqn:fp2}
   |f_{\rm p}|^2
 = \frac{1}{\rho_0}
   \bbrac{\frac{\pi}{4 \sqrt{2}} \afrac{1}{\calE A^2 \sqrt{\mN}}}
   \abrac{\mchi+\mN} \bbrac{\frac{2 \Qmin^{1/2} r(\Qmin)}{\FQmin}+I_0}\, .
\eeq
Note that the factor $\calE$ appearing in the denominator is the exposure of the
experiment, which is dimensionless in natural units. It relates the actual
counting rate to the normalized rate of Eq.(\ref{eqn:dRdQ}). The WIMP mass
$\mchi$ on the right--hand side can be estimated by our method described in
Ref.~\cite{DMDDmchi} using data from two experiments. $r(\Qmin)$ and $I_0$ can
be estimated from one of the two data sets used for determining $\mchi$ or
from a third experiment. Recall that, due to the degeneracy between the local
WIMP density $\rho_0$ and the WIMP--nucleus cross section $\sigma_0$, one {\em can
  not} estimate each one of them without using some assumptions. Hence, by
using Eq.(\ref{eqn:fp2}), one will need to accept an assumption of the local
WIMP density $\rho_0$.\footnote{The most commonly used value for the local WIMP
  density is $\rho_0 \approx 0.3~{\rm GeV/cm^3}$. However, an uncertainty of
  a factor of $\sim$ 2 has been usually adopted: $\rho_0 = 0.2 - 0.8~{\rm
    GeV/cm^3}$.}

In Figs.~1, we show the reconstructed spin--independent WIMP--nucleon coupling
$|f_{\rm p}|_{\rm rec}^2$ as a function of the {\em input} WIMP mass
$m_{\chi,{\rm in}}$. Following our work on determination of the WIMP mass
\cite{DMDDmchi}, $\rmXA{Ge}{76}$ and $\rmXA{Si}{28}$ have been chosen as two
targets for estimating $\mchi$. In order to avoid calculating correlations
between $\mchi$ and $I_0$, a second {\em independent} data set with
$\rmXA{Ge}{76}$ (left frame) or $\rmXA{Si}{28}$ (right frame) has been chosen
for estimating $I_0$.
\begin{figure}[p]
\begin{center}
\includegraphics[width=6.5cm]{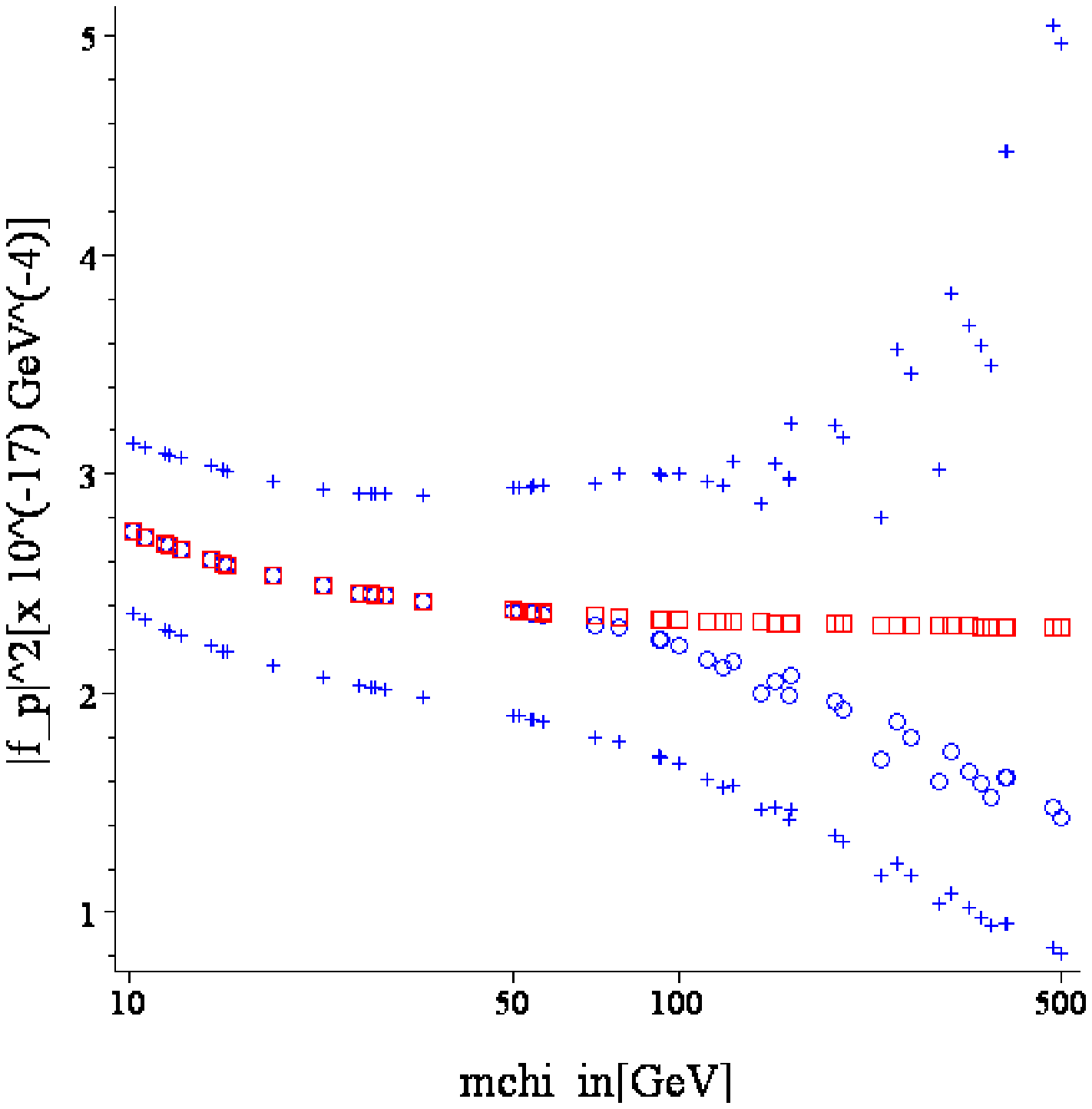} \hspace{1.5cm}
\includegraphics[width=6.5cm]{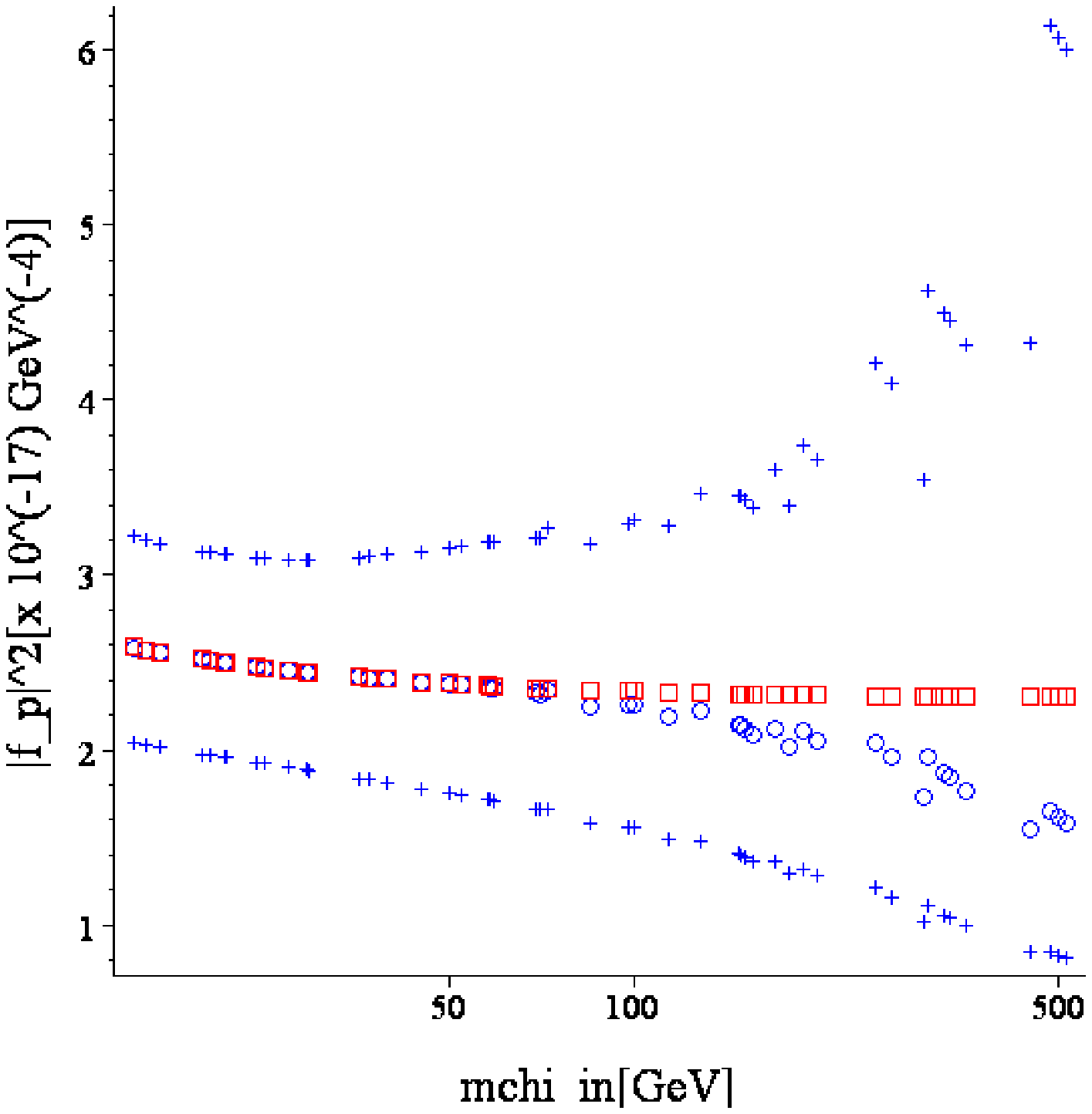} \\
 (a) $\rmXA{Ge}{76}$ \hspace{6.6cm}
 (b) $\rmXA{Si}{28}$
\end{center}
\caption{
  The reconstructed spin--independent WIMP--nucleon coupling $|f_{\rm p}|_{\rm
    rec}^2$ as a function of the {\em input} WIMP mass $m_{\chi,{\rm in}}$.
  The open (red) squares indicate the input WIMP masses and the true values of
  the SI WIMP--nucleon couplings.  The open (blue) circles and the (blue)
  crosses indicate the reconstructed SI WIMP--nucleon couplings and the
  1$\sigma$ statistical errors.  The theoretical predicted recoil spectrum for
  the shifted Maxwellian velocity distribution function \cite{SUSYDM},
  \cite{DMDDf1v} with Woods-Saxon elastic form factor \cite{Engel91},
  \cite{SUSYDM} ($v_0 = 220$ km/s, $v_e = 231$ km/s) have been used.  The
  WIMP--nucleon cross section has been set to be \mbox{$10^{-8}$ pb}.
  $\rmXA{Ge}{76}$ and $\rmXA{Si}{28}$ have been chosen as two targets for
  estimating $\mchi$.  A second (independent) data set with $\rmXA{Ge}{76}$
  (left frame) or $\rmXA{Si}{28}$ (right frame) have been chosen as the third
  nucleus for estimating $I_0$.  Each experimental data set has 50 events
  under the maximal cut--off energy $\Qmax$ chosen as 100 keV.}
\end{figure}
\begin{figure}[p]
\begin{center}
\includegraphics[width=6.5cm]{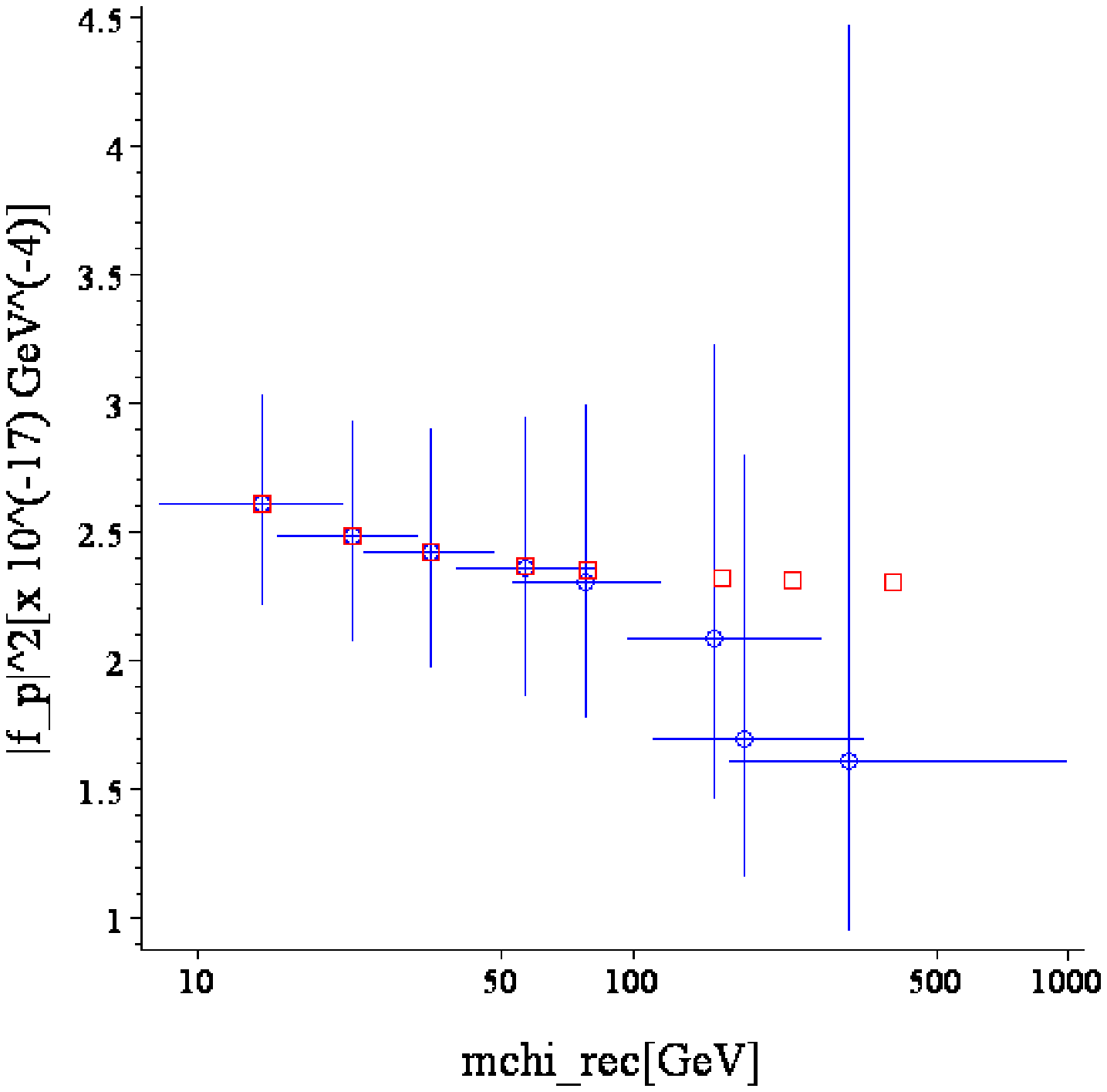} \hspace{1.5cm}
\includegraphics[width=6.5cm]{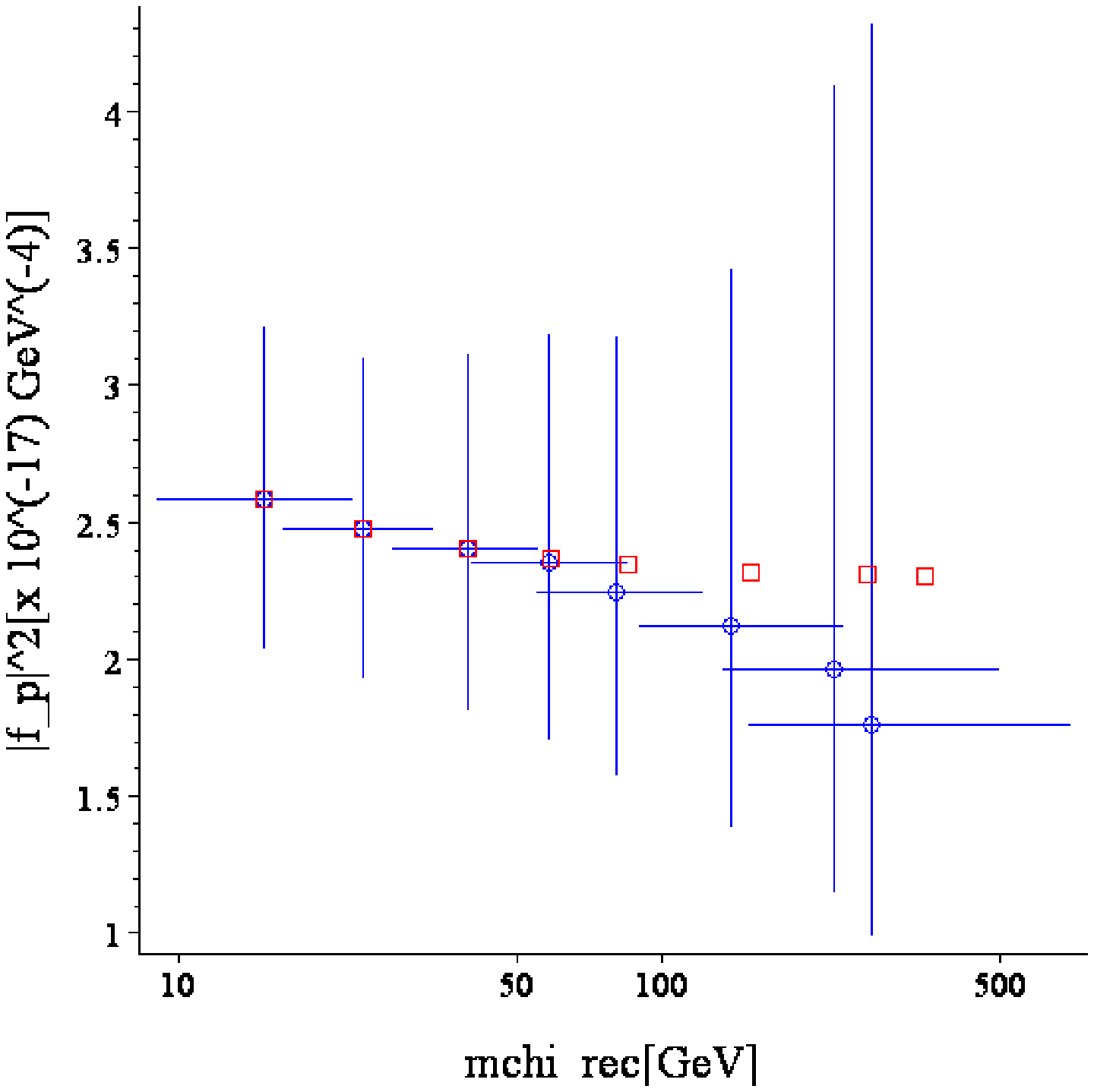} \\
 (a) $\rmXA{Ge}{76}$ \hspace{6.6cm}
 (b) $\rmXA{Si}{28}$
\end{center}
\caption{
  The reconstructed spin--independent WIMP--nucleon coupling $|f_{\rm p}|_{\rm
    rec}^2$ and the {\em reconstructed} WIMP mass $m_{\chi,{\rm rec}}$ on the
  cross section (coupling) v.s. WIMP mass plane.  The open (red) squares
  indicate the input WIMP masses and the true values of the SI WIMP--nucleon
  couplings.  The open (blue) circles indicate the reconstructed WIMP masses
  and the reconstructed SI couplings on nucleon.  The horizontal and vertical
  solid (blue) lines show the 1$\sigma$ statistical errors on $m_{\chi,{\rm
      rec}}$ and $|f_{\rm p}|_{\rm rec}^2$, respectively.  Parameters as in
  Fig.~1.}
\end{figure}

It can be seen in Figs.~1 that the reconstructed $|f_{\rm p}|_{\rm rec}^2$
values are {\em underestimated} for WIMP masses $\gsim~100$ GeV. For Ge this
deviation is larger than for Si. This systematic deviation is caused by the
underestimate of $I_0$. It is worse for heavier nuclei, where events with
higher recoil energies can contribute more. This deviation of $I_0$ could be
reduced by extending the maximal cut--off energy $\Qmax$ to higher energy
range since the kinematic maximal cut--off energy is larger for heavier WIMP
masses. On the other hand, although we used the same event number for both
experiments the statistical error on $I_0$ estimated with Si is larger than
that with Ge.

Nevertheless, it can be seen in Figs.~1 that, first, in spite of this
systematic deviation, the true value of $|f_{\rm p}|_{\rm rec}^2$ always lies
within the 1$\sigma$ error interval.  Second, for a WIMP mass of 100 GeV, one
could in principle already estimate the SI WIMP--nucleon coupling with a
statistical uncertainty of only $\sim$ 15\% with just 50 events from each
experiment.

Combining the estimate for $|f_{\rm p}|^2$ with that for $\mchi$, in Figs.~2
we show the reconstructed SI coupling $|f_{\rm p}|_{\rm rec}^2$ and the {\em
  reconstructed} WIMP mass $m_{\chi,{\rm rec}}$ on the cross section
(coupling) v.s. WIMP mass plane. We emphasize that by our methods described in
Ref.~\cite{DMDDmchi} and here, one can estimate $\mchi$ and $|f_{\rm p}|^2$
{\em separately} {\em without} any assumption for the WIMP velocity
distribution.

\section{Conclusions}

In this paper we have extended our method for determining the WIMP mass
\cite{DMDDmchi} to estimate the spin--independent WIMP--nucleon coupling from
the elastic WIMP--nucleus scattering experiments.  This method is independent
of the velocity distribution of halo WIMPs as well as (practically) of the as
yet unknown WIMP mass.  The only information needed is the measured recoil
energies from at least two experiments with different target nuclei and the
local Dark Matter density as the unique assumption.

These information combined with the reconstructed WIMP mass will allow us not
only to constrain the parameter space in different extensions of the Standard
Model of particle physics, but also to identify WIMPs among new particles
produced at colliders (hopefully in the near future).  Once one is confident
of this identification, one can use further collider measurements of the mass
and couplings of WIMPs.  Together with the reconstruction of the velocity
distribution of halo WIMPs \cite{DMDDf1v}, this will then yield a new
determination of the local WIMP density.  On the other hand, knowledge of the
WIMP couplings will also permit prediction of the WIMP annihilation cross
section.  Together with information on the WIMP density, this will allow to
predict the event rate in the indirect Dark Matter detection as well as to
test our understanding of the early Universe.

\subsubsection*{Acknowledgments}
This work was partially supported by the Marie Curie Training
Research Network ``UniverseNet'' under contract no.~MRTN-CT-2006-035863, by
the European Network of Theoretical Astroparticle Physics ENTApP ILIAS/N6
under contract no.~RII3-CT-2004-506222, as well as by the BK21 Frontier
Physics Research Division under project no.~BA06A1102 of Korea Research
Foundation.

\end{document}